\documentclass[12pt]{iopart}
\usepackage{graphicx}

\begin{document}

\title[Higher order corrections for shallow-water solitary waves]
{Higher order corrections for shallow-water solitary waves: elementary derivation and experiments}

\author{G\'{a}bor B Hal\'{a}sz$^{1,2}$}

\address{$^1$Trinity College, University of Cambridge, Trinity
Street, Cambridge CB2 1TQ, UK \\
$^2$von K\'{a}rm\'{a}n Laboratory for Environmental Flows,
E\"{o}tv\"{o}s University, P\'{a}zm\'{a}ny P\'{e}ter s\'{e}t\'{a}ny
1/A, H-1117 Budapest, Hungary}

\ead{gh315@cam.ac.uk}


\begin{abstract}

We present an elementary method to obtain the equations of the
shallow-water solitary waves in different orders of approximation.
The first two of these equations are solved to get the shapes and
propagation velocities of the corresponding solitary waves. The
first-order equation is shown to be equivalent to the Korteweg$-$de
Vries (KdV) equation, while the second-order equation is solved
numerically. The propagation velocity found for the solitary waves
of the second-order equation coincides with a known expression, but
it is obtained in a simpler way. By measuring the propagation
velocity of solitary waves in the laboratory, we demonstrate that
the second-order theory gives a considerably improved fit to
experimental results.

\end{abstract}


\maketitle


\section{Introduction} \label{sect-intro}

Solitary waves propagating on the surface of fluids have been
extensively studied since John Scott Russell discovered them in 1834
\cite{Drazin}. They occur naturally in the form of tidal bores, but
they can be generated in a laboratory as well. The most striking
property of these solitary waves is their well-distinguished shape
which they maintain during propagation; they can travel extremely
long distances without considerable dispersion or dissipation.

The first theory to successfully explain the existence of solitary
waves was developed by Korteweg and de Vries \cite{KdV}. It is based
on the shallow-water theory of ideal fluids, which predicts linear
waves in the small amplitude limit; these waves are slightly
dispersive. However, nonlinearity is also present due to the
convective term appearing in the Euler equation. According to the
Korteweg$-$de Vries (KdV) theory, solitary waves emerge as a balance
between nonlinearity and dispersion. The shape of the waves and
their propagation velocity can be obtained from the exactly solvable
KdV equation.

The KdV theory can be considered as a first-order approximation to
solitary waves. Although it explains their unusual properties and
agrees well with experiments for small amplitude solitary waves,
further refinements to the theory are possible. By using the
systematic expansion method developed by Friedrichs
\cite{Friedrichs}, the second-order approximation to the solitary
waves was found by Laitone \cite{Laitone}. Later on, Fenton extended
the method to nine orders \cite{Fenton}, while Schwartz reached the
70th-order approximation with the aid of computers \cite{Schwartz}.

The derivations of the KdV equation or any higher order
approximations are absent from many standard textbooks
\cite{Landau,Kundu}, while others giving more complete account on
the topic use involved mathematical techniques \cite{Drazin,Ludu}.
In this paper, we present an alternative method for treating
shallow-water solitary waves. This method is mathematically simpler
and physically more intuitive; it can be presented in any
undergraduate course. Based merely on the conditions of
incompressibility and irrotational flow, we use Bernoulli's law to
obtain equations that describe the shape of solitary waves in the
different order approximations mentioned above. The first two
equations are solved to find approximate shapes and two expressions
for the propagation velocities. We test the validity of these
expressions by comparing them to large amplitude solitary waves in
the laboratory, and find that the second-order approximation gives
much better correlation with the experimental results.

\begin{figure}[t!]
\centering
\includegraphics[width=9cm]{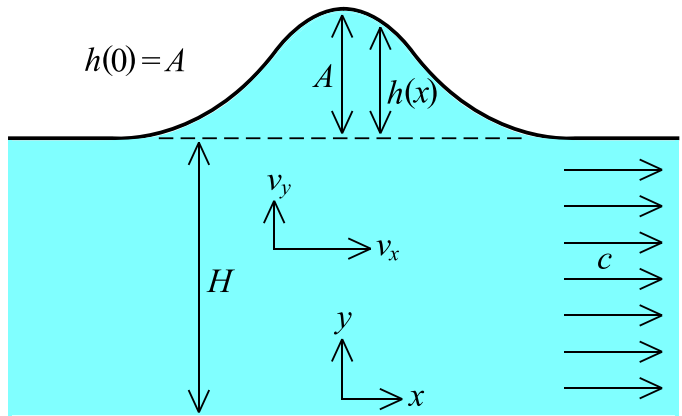}
\caption{An illustration describing the stationary flow of a
solitary wave in a co-moving frame. \label{figure1}}
\end{figure}

\section{Basic equations} \label{sect-basic}

Let us consider a solitary wave propagating to the left along the
$-x$ direction with velocity $c$ in an unbounded fluid of ambient
height $H$. Examining it from a co-moving system, the stationary
flow described in Figure \ref{figure1} is observed. The origin of
the coordinate system is placed to the bottom of the fluid under the
peak of the solitary wave. The horizontal ($v_x$) and vertical
($v_y$) components of the flow velocity are functions of the
coordinates $x$ and $y$, while $h(x)$ denotes the excess fluid
height compared to $H$. The amplitude of the solitary wave is
defined as the value $A = h(0)$. In the limiting case of  $x
\rightarrow \pm \infty$, it is obvious that $v_x(x,y) \rightarrow
c$, $v_y(x,y) \rightarrow 0$ and $h(x) \rightarrow 0$.

For many liquids, in particular water, the effect of viscosity can
be neglected, therefore the problem can be described by the Euler
equation. Instead of trying to solve it directly, we first recite
some important properties of the flow to be examined: first, water
is practically incompressible, hence
\begin{equation}
\nabla \underline{v} = 0. \label{cond-div}
\end{equation}
As a consequence, the density $\rho$ of the liquid is constant. It
can be assumed by most wave phenomena that the flow is irrotational
so that
\begin{equation}
\nabla \times \underline{v} = 0. \label{cond-rot}
\end{equation}
According to the standard boundary conditions, the vertical velocity
must vanish at the bottom of the liquid:
\begin{equation}
v_y = 0 \quad \textrm{for} \quad y = 0. \label{cond-bottom}
\end{equation}
It is also clear that the material flux through the full depth must
be the same for all vertical cross-sections taken at any $x$. The
constant value of this flux can be obtained by calculating it for
the limiting case of $x \rightarrow \pm \infty$ such that
\begin{equation}
\int_{y=0}^{H+h(x)} v_x(x,y) dy = cH. \label{cond-rate}
\end{equation}
The last important property of the flow is that the liquid pressure
equals the pressure of air on the surface:
\begin{equation}
p = p_0 \quad \textrm{for} \quad y = H+h(x). \label{cond-top}
\end{equation}
Finding a flow that satisfies conditions
(\ref{cond-div})$-$(\ref{cond-top}) is much easier than solving the
Euler equation directly.

\section{An iteration scheme} \label{sect-iter}

In this section, we present a method for obtaining an ordinary
differential equation for $h(x)$ in an iterative sequence of steps.
Let us first examine the flow with the velocity components
\begin{equation}
v_x(x,y) = \frac{cH}{H+h(x)} \quad \textrm{and} \quad v_y(x,y) = 0.
\label{comp-xy}
\end{equation}
It is clear that the conditions (\ref{cond-rot})$-$(\ref{cond-rate})
are fulfilled; however, the divergence of the velocity does not
vanish. To satisfy condition (\ref{cond-div}), we add a new term to
the vertical velocity $v_y$. The new term $\Delta v_y$ can be
obtained by elementary methods: differentiation of $v_x$ with
respect to the coordinate $x$ and integration with respect to the
single variable $y$. An arbitrary function of $x$ appears after the
integration which can be chosen to fulfil condition
(\ref{cond-bottom}). The new term obviously does not contribute to
the flux, therefore the flow with these modified components
satisfies conditions (\ref{cond-div}), (\ref{cond-bottom}) and
(\ref{cond-rate}).

For the new flow, however, the rotation of the velocity does not
disappear; to satisfy condition (\ref{cond-rot}), we now add a new
term $\Delta v_x$ to the horizontal velocity $v_x$. The method for
obtaining $\Delta v_x$ is nearly the same as above; after the
differentiation of $v_y$ with respect to $x$ and the integration
with respect to $y$ an arbitrary function of $x$ appears. This can
be chosen to make the flux resulting from the new term vanish. The
vertical velocity at the bottom of the water obviously remains zero,
hence the flow with the new term satisfies the conditions
(\ref{cond-rot})$-$(\ref{cond-rate}).

\begin{figure}[h]
\centering
\includegraphics[width=10cm]{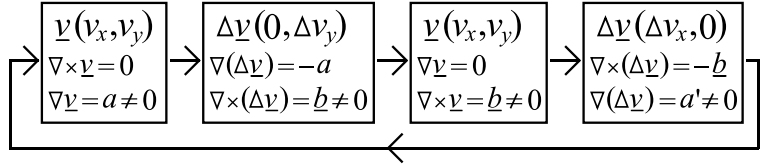}
\caption{A flow diagram summarizing one iteration cycle, as
described in the text. \label{figure2}}
\end{figure}

At this point, the situation is formally the same as at the
beginning: the divergence of the flow is non-zero again. It can be
assumed, however, that after these steps we are closer to the exact
flow solution. Once again, a new term must be added to the vertical
velocity and another one to the horizontal velocity as mentioned
above. By following this iteration scheme and repeating the steps
shown in Figure \ref{figure2}, we obtain infinitely long expressions
for both components of the velocity. Assuming that this method is
convergent, the new terms become less and less significant. The
remaining divergence and rotation after all the steps therefore
approach zero and the limiting flow satisfies conditions
(\ref{cond-div})$-$(\ref{cond-rate}).

It is, however, still left to examine whether condition
(\ref{cond-top}) is fulfilled. Since viscosity can be neglected and
the flow is irrotational, Bernoulli's law holds between two points
of the water surface; one of these points can be chosen to be
infinitely far away, while the other is arbitrary with its
horizontal coordinate $x$:
\begin{equation}
\fl p_0 + \rho gH + \frac{1}{2} \rho c^2 = p_0 + \rho g \big{(}
H+h(x) \big{)} + \frac{1}{2} \rho \Big{[} v_x^2 \big{(} x,H+h(x)
\big{)} + v_y^2 \big{(} x,H+h(x) \big{)} \Big{]}. \label{eqn-bern1}
\end{equation}
After rearranging this equation, both the air pressure $p_0$ and the
density $\rho$ of water cancel out, hence we obtain
\begin{equation}
1 - \frac{2gh(x)}{c^2} = \frac{v_x^2 (x,H+h(x))}{c^2} + \frac{v_y^2
(x,H+h(x))}{c^2}. \label{eqn-bern2}
\end{equation}
This equation determines the shape $h(x)$ of the surface. The
velocity components are given by the infinitely long expressions
obtained from the iteration scheme of Figure \ref{figure2}. Equation
(\ref{eqn-bern2}) is thus an infinitely long ordinary differential
equation, which is impossible to treat without approximations.

\section{Order of magnitude estimates} \label{sect-magn}

By substituting in principle the infinitely long expressions of
$v_x(x,H+h(x))$ and $v_y(x,H+h(x))$ into equation (\ref{eqn-bern2})
we obtain an equation containing an infinite number of complicated
terms. These terms can, however, be expanded into series with
respect to dimensionless quantities much less than unity. Assuming a
long solitary wave of small amplitude, these quantities are
\begin{equation}
\frac{h(x)}{H} \ll 1 \quad \textrm{and} \quad H h''(x) \ll 1,
\label{term-small1}
\end{equation}
for example. The derivatives of $h(x)$ with respect to $x$ are
denoted by primes. After the expansion the equation is still
infinitely long, but its terms are simpler and can more easily be
classified. Equation (\ref{eqn-bern2}) can formally be written as
\begin{equation}
\frac{2gh(x)}{c^2} - \frac{2h(x)}{H} + \sum_{n=1}^{\infty} T_n = 0,
\label{eqn-Tn}
\end{equation}
where all terms denoted by $T_n$ take the similar form: they are the
products of the function $h(x)$ and its different derivatives
multiplied or divided by the appropriate power of $H$ to keep them
dimensionless. Some of the terms are nonlinear, while those
containing derivatives are responsible for dispersion. Examples of
such terms are
\begin{equation}
\frac{h(x)^2}{H^2} \quad \textrm{and} \quad H \cdot h''(x).
\label{term-h1}
\end{equation}
Both nonlinearity and dispersion can occur in more complex terms,
such as
\begin{equation}
H^2 \cdot h'(x) \cdot h'''(x). \label{term-h2}
\end{equation}
To estimate the magnitudes of these different terms, let us write
the shape of the solitary wave as
\begin{equation}
h(x) = A \cdot f(kx) \quad \textrm{where} \quad f(0) = 1.
\label{shape-gen}
\end{equation}
The new quantity $k$ is an 'effective wave number', which is
inversely proportional to the horizontal extension of the solitary
wave, while $f$ denotes an appropriately smooth unknown function of
the dimensionless product $kx$. The magnitude of the derivatives of
$h(x)$ can be estimated via equation (\ref{shape-gen}). For example,
the magnitudes of the terms in (\ref{term-h1}) are
\begin{equation}
\frac{h(x)^2}{H^2} \sim \frac{A^2}{H^2} \quad \textrm{and} \quad H
\cdot h''(x) \sim H \cdot Ak^2, \label{term-h1-magn}
\end{equation}
respectively. It is assumed that the amplitude $A$ is much smaller,
while the length $\sim 1/k$ is much larger than the initial water
height $H$:
\begin{equation}
\frac{A}{H} \ll 1 \quad \textrm{and} \quad k H \ll 1,
\label{term-small2}
\end{equation}
which are equivalent to the relations in (\ref{term-small1}).

Solitary waves emerge as a balance between nonlinearity and
dispersion; in the most simplistic case, it is enough to keep the
largest nonlinear and the largest dispersive term. These are the
terms in (\ref{term-h1}), therefore in this case they must have the
same order of magnitude: $A^2/H^2 \sim H Ak^2$, i.e.
\begin{equation}
\frac{A}{H} \sim k^2 H^2. \label{term-comp}
\end{equation}
This relation determines the relative magnitudes of the small
quantities in (\ref{term-small2}). In the following section, we
derive the simplest approximation, which corresponds to a case where
all terms $T_n$ in equation (\ref{eqn-Tn}) are neglected except for
the terms in (\ref{term-h1}). The solution can be checked to be in
agreement with relation (\ref{term-comp}). It is natural to expect
that the same relation also holds for the more accurate
approximations since keeping smaller terms does not essentially
change the order of magnitude relationships.\\

\section{First- and second-order equations} \label{sect-eqn}

In order to derive the approximate equations, let us return to
equation (\ref{comp-xy}) and implement the method described in the
previous two sections. First of all, we divide the velocity
components by $c$, and expand the component $v_x$ into a Taylor
series. If we keep terms up to the order of $A^3/H^3$, we obtain
\begin{equation}
\frac{v_x}{c} = 1 - \frac{h(x)}{H} + \frac{h(x)^2}{H^2} -
\frac{h(x)^3}{H^3} \quad \textrm{and} \quad \frac{v_y}{c} = 0.
\label{comp-xy1}
\end{equation}
Now we add a new term $\Delta v_y$ to the vertical velocity and
hence satisfy condition (\ref{cond-div}):
\begin{equation}
\frac{\partial v_x}{\partial x} + \frac{\partial (\Delta
v_y)}{\partial y} = 0. \label{comp-dy2-eqn}
\end{equation}
By keeping terms up to the order of $A^3/H^3$, we obtain
\begin{equation}
\frac{\Delta v_y}{c} = - \frac{1}{c} \int \frac{\partial
v_x}{\partial x} dy = \frac{h'(x)}{H} \bigg{(} 1 - \frac{2h(x)}{H}
\bigg{)} y + g(x). \label{comp-dy2}
\end{equation}
The arbitrary function $g(x)$ should be chosen to be zero in order
to fulfil condition (\ref{cond-bottom}). Next, we add a new term
$\Delta v_x$ to the horizontal velocity to cancel out the rotation
of the flow component (\ref{comp-dy2}):
\begin{equation}
\frac{\partial (\Delta v_y)}{\partial x} - \frac{\partial (\Delta
v_x)}{\partial y} = 0. \label{comp-dx2-eqn}
\end{equation}
After the integration with respect to $y$ there remains an arbitrary
function $q$ of $x$:
\begin{equation}
\frac{\Delta v_x}{c} = \frac{1}{c} \int \frac{\partial (\Delta
v_y)}{\partial x} dy = \frac{1}{2H^2} \Big{[} \big{(} H-2h(x)
\big{)} h''(x) - 2 h'(x)^2 \Big{]} y^2 + q(x). \label{comp-dx2}
\end{equation}
The function $q(x)$ should be chosen to make the flux resulting from
the component (\ref{comp-dx2}) equal to zero:
\begin{equation}
\int_{y=0}^{H+h(x)} \frac{1}{2H^2} \Big{[} \big{(} H-2h(x) \big{)}
h''(x) - 2 h'(x)^2 \Big{]} y^2 dy + q(x) \big{(} H+h(x) \big{)} = 0.
\label{comp-dx2-rate}
\end{equation}
Hence the velocity components after the first iteration cycle are
given by
\begin{eqnarray}
\fl \frac{v_x}{c} = \bigg{(} 1 - \frac{h(x)}{H} + \frac{h(x)^2}{H^2}
- \frac{h(x)^3}{H^3} \bigg{)} \nonumber \\
+ \frac{1}{2H^2} \Big{[} \big{(} H-2h(x) \big{)} h''(x) - 2 h'(x)^2
\Big{]} \bigg{(} y^2 - \frac{1}{3} \big{(} H+h(x) \big{)}^2 \bigg{)}
\label{comp-x2}
\end{eqnarray}
and
\begin{equation}
\frac{v_y}{c} = \frac{h'(x)}{H} \bigg{(} 1 - \frac{2h(x)}{H}
\bigg{)} y. \label{comp-y2}
\end{equation}
The substitution of these components into equation (\ref{eqn-bern2})
yields
\begin{eqnarray}
\fl 1 - \frac{2gh(x)}{c^2} = \bigg{[} \bigg{(} 1 - \frac{h(x)}{H} +
\frac{h(x)^2}{H^2} - \frac{h(x)^3}{H^3} \bigg{)} \nonumber \\
+ \frac{1}{3} \bigg{(} 1 + \frac{h(x)}{H} \bigg{)}^2 \Big{[} \big{(}
H-2h(x) \big{)} h''(x) - 2 h'(x)^2 \Big{]} \bigg{]}^2 + h'(x)^2
\label{eqn-bern3}
\end{eqnarray}
if we keep terms up to the order of $A^3/H^3$. The magnitude of the
different terms can be estimated via equations (\ref{shape-gen}) and
(\ref{term-comp}), similarly as for the terms in
(\ref{term-h1-magn}):
\begin{equation}
\frac{h(x)}{H} \sim \frac{A}{H}, \quad \frac{h(x)^2}{H^2} \sim H
\cdot h''(x) \sim \frac{A^2}{H^2} \label{magn-term1}
\end{equation}
and
\begin{equation}
\frac{h(x)^3}{H^3} \sim h(x) \cdot h''(x) \sim h'(x)^2 \sim
\frac{A^3}{H^3}. \label{magn-term2}
\end{equation}
Rearranging the right-hand side of equation (\ref{eqn-bern3}) and
keeping the terms up to the order of $A^3/H^3$ leads to the form
(\ref{eqn-Tn}):
\begin{equation}
\frac{2gh(x)}{c^2} - \frac{2h(x)}{H} + T_1 + T_2 = 0,
\label{eqn-T12}
\end{equation}
where
\begin{equation}
T_1 = \frac{3h(x)^2}{H^2} + \frac{2}{3} H \cdot h''(x)
\label{term-T1}
\end{equation}
and
\begin{equation}
T_2 = - \frac{4h(x)^3}{H^3} - \frac{2}{3} h(x) \cdot h''(x) -
\frac{1}{3} h'(x)^2 \label{term-T2}
\end{equation}
are on the order of $A^2/H^2$ and $A^3/H^3$, respectively.

The simplest approximation is to neglect both $T_1$ and $T_2$ in
equation (\ref{eqn-T12}). In this case, we obtain the linear
equation
\begin{equation}
\bigg{(} \frac{2g}{c^2} - \frac{2}{H} \bigg{)} h(x)= 0,
\label{eqn-order0}
\end{equation}
which describes the shallow-water linear waves of arbitrary shape.
These non-dispersive waves propagate with the well-known velocity of
\begin{equation}
c = c_0 = \sqrt{gH}, \label{vel-order0}
\end{equation}
which converts equation (\ref{eqn-order0}) into an identity for any
$h(x)$.

Nonlinear and dispersive terms are, however, both necessary for a
solitary wave solution. The simplest approximation describing such
solutions can be obtained by keeping $T_1 \sim A^2/H^2$ and still
neglecting $T_2 \sim A^3/H^3$ in equation (\ref{eqn-T12}):
\begin{equation}
\frac{2gh(x)}{c^2} - \frac{2h(x)}{H} + \frac{3h(x)^2}{H^2} +
\frac{2}{3} H \cdot h''(x) = 0. \label{eqn-order1}
\end{equation}
The first-order approximate equation (\ref{eqn-order1}) is, to the
given approximation, equivalent to the KdV equation, which is
generally used to describe solitary waves \cite{KdV}. Since relation
(\ref{vel-order0}) is a zeroth-order solution for the propagation
velocity $c$, it is worth introducing the small dimensionless
quantity
\begin{equation}
\epsilon = 1 - \frac{gH}{c^2} \ll 1 \label{eps-def}
\end{equation}
and approximating the first two terms in equation (\ref{eqn-order1})
as
\begin{equation}
\fl \bigg{(} \frac{2gH}{c^2} - 2 \bigg{)} \frac{h(x)}{H} = \bigg{[}
4 - 4 \bigg{(} 1 + \frac{\epsilon}{2} \bigg{)} \bigg{]}
\frac{h(x)}{H} \approx \bigg{[} 4 - \frac{4}{\sqrt{1-\epsilon}}
\bigg{]} \frac{h(x)}{H} = \bigg{(} 4 - \frac{4c}{\sqrt{gH}} \bigg{)}
\frac{h(x)}{H}. \label{eps-approx}
\end{equation}
The approximation $(1+\epsilon)^\alpha \approx 1 + \alpha \epsilon$
appears in the second step, which is valid to first order in
$\epsilon$. After differentiation and multiplication by $H/4$ we
obtain
\begin{equation}
h'(x) - \frac{c}{\sqrt{gH}} h'(x) + \frac{3}{2H} h(x) \cdot h'(x) +
\frac{1}{6} H^2 h'''(x) = 0 \label{eqn-order1-mod}
\end{equation}
from equation (\ref{eqn-order1}). This ordinary differential
equation is then transformed into a partial one with the independent
variables $x$ and $t$. Derivatives in all terms of equation
(\ref{eqn-order1-mod}) must be substituted with a partial derivative
either with respect to $x$ or with respect to $t$. In the latter
case, however, the term also must be divided by $c$ so that equation
(\ref{eqn-order1-mod}) can be restored by seeking the solution in
the special form $h(x+ct)$. Since the propagation velocity $c$ is
different for every special wave solution, it must not explicitly
appear in an equation describing such a wide range of phenomena.
Only the second term in equation (\ref{eqn-order1-mod}) should
therefore contain a time derivative. By leaving spatial derivatives
in all other terms we obtain
\begin{equation}
\frac{\partial h}{\partial x} - \frac{1}{\sqrt{gH}} \cdot
\frac{\partial h}{\partial t} + \frac{3}{2H} h \cdot \frac{\partial
h}{\partial x} + \frac{1}{6} H^2 \frac{\partial^3 h}{\partial x^3} =
0, \label{eqn-order1-KdV}
\end{equation}
the KdV equation describing waves propagating to the left.

It can be verified that further steps after the first iteration
cycle do not result in new terms with magnitudes up to the order of
$A^2/H^2$. For the second-order approximation, however, we must
collect all terms with magnitudes up to $A^3/H^3$; this requires one
more iteration cycle. If we keep terms up to the order of $A^3/H^3$,
the new term $\Delta v_y'$ added to the vertical velocity is given
by
\begin{equation}
\frac{\Delta v_y'}{c} = - \frac{1}{c} \int \frac{\partial (\Delta
v_x)}{\partial x} dy = - \frac{1}{6H} h'''(x) \Big{[} y^3 - y
\big{(} H+h(x) \big{)}^2 \Big{]} + g(x). \label{comp-dy3}
\end{equation}
The arbitrary function $g(x)$ is once again chosen to be zero so
that the flow fulfils condition (\ref{cond-bottom}). By still
keeping terms up to $A^3/H^3$, the new term $\Delta v_x'$ added to
the horizontal velocity reads as
\begin{equation}
\frac{\Delta v_x'}{c} = \frac{1}{c} \int \frac{\partial (\Delta
v_y')}{\partial x} dy = -\frac{1}{24H} h^{(4)}(x) \Big{[} y^4 - 2y^2
\big{(} H+h(x) \big{)}^2 \Big{]} + r(x), \label{comp-dx3}
\end{equation}
where the function $r(x)$ is chosen to make the flux resulting from
the flow component (\ref{comp-dx3}) vanish:
\begin{equation}
\int_{y=0}^{H+h(x)} - \frac{1}{24H} h^{(4)}(x) \Big{[} y^4 - 2y^2
\big{(} H+h(x) \big{)}^2 \Big{]} dy + r(x) \big{(} H+h(x) \big{)} =
0. \label{comp-dx3-rate}
\end{equation}
Hence the new term $\Delta v_x'$ is given by
\begin{equation}
\frac{\Delta v_x'}{c} = -\frac{h^{(4)}(x)}{360H} \Big{[} 15y^4 -
30y^2 \big{(} H+h(x) \big{)}^2 + 7 \big{(} H+h(x) \big{)}^4
\Big{]}.\label{comp-dx3-final}
\end{equation}
Similarly as after the first iteration cycle, the velocity
components of the net flow obtained from expressions
(\ref{comp-x2}), (\ref{comp-y2}), (\ref{comp-dy3}) and
(\ref{comp-dx3-final}) are substituted into equation
(\ref{eqn-bern2}), which is then rearranged to the form
(\ref{eqn-Tn}). By keeping terms up to the order of $A^3/H^3$, the
three terms of $T_2$ in equation (\ref{term-T2}) appear with respect
to equation (\ref{eqn-order1}). One new term, $2H^3 h^{(4)}(x)/45$,
also results from the component (\ref{comp-dx3-final}) of the second
iteration cycle. The second-order approximate equation thus takes
the form
\begin{equation}
\fl \frac{2gh(x)}{c^2} - \frac{2h(x)}{H} + \frac{3h(x)^2}{H^2} +
\frac{2}{3} H \cdot h''(x) - \frac{4h(x)^3}{H^3} - \frac{2}{3} h(x)
\cdot h''(x) - \frac{1}{3} h'(x)^2 + \frac{2}{45} H^3 h^{(4)}(x) =
0. \label{eqn-order2}
\end{equation}
By continuing the same method it is possible to obtain more accurate
(higher order) approximations without too much difficulty. The
calculations, however, become far more complicated, hence these
further equations are beyond the scope of this paper.

\section{Solitary wave solutions} \label{sect-sol}

\begin{figure}[t!]
\centering
\includegraphics[width=6cm]{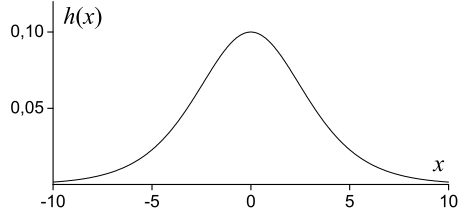}
\caption{Shape of the solitary wave of the first-order equation
(\ref{eqn-order1}) with water height $H$ = 1 and amplitude $A$ =
0.1. \label{figure3}}
\end{figure}

In this section, we first recover the well-known properties of the
KdV solitary waves from equation (\ref{eqn-order1}), then find a
solitary wave solution for the second-order approximate equation
(\ref{eqn-order2}). By using abbreviation (\ref{eps-def}), equation
(\ref{eqn-order1}) reads as
\begin{equation}
-2 \epsilon \frac{h(x)}{H} + 3 \frac{h(x)^2}{H^2} + \frac{2}{3} H
\cdot h''(x) = 0. \label{eqn-order1-eps}
\end{equation}
After a multiplication by $h'(x)$ it can be integrated to give
\begin{equation}
-\epsilon \frac{h(x)^2}{H} + \frac{h(x)^3}{H^2} + \frac{1}{3} H
\cdot h'(x)^2 = B. \label{eqn-order1-int1}
\end{equation}
In solitary waves $h'(x) \rightarrow 0$ for $h(x) \rightarrow 0$,
hence the integration constant $B$ is zero. By expressing $h'(x)$
from equation (\ref{eqn-order1-int1}) we obtain the separable
differential equation
\begin{equation}
h'(x) = \pm \sqrt{\frac{3}{H^3}} \cdot h(x) \sqrt{\epsilon H -
h(x)}. \label{eqn-order1-int2}
\end{equation}
The solution for equation (\ref{eqn-order1-int2}) can be obtained by
elementary methods:
\begin{equation}
h(x) = \epsilon H \cdot \cosh^{-2} \big{(} k(x-C) \big{)},
\label{shape-order1}
\end{equation}
where $k$ is the 'effective wave number' given by
\begin{equation}
k = \sqrt{\frac{3\epsilon}{4H^2}}. \label{k-def}
\end{equation}
The integration constant $C$ is only responsible for shifting the
wave along the $x$-axis, therefore it can be chosen to be zero
without any restriction. By comparing equations (\ref{shape-gen}),
(\ref{shape-order1}) and (\ref{k-def}) we obtain
\begin{equation}
\epsilon = \frac{A}{H} = \frac{4}{3} k^2 H^2 \label{eps-order1}
\end{equation}
between the amplitude and the length of the solitary wave, clearly
verifying the validity of relation (\ref{term-comp}). The
propagation velocity $c$ can be expressed from equations
(\ref{eps-def}) and (\ref{eps-order1}) by means of the following
approximation:
\begin{equation}
c = \frac{\sqrt{gH}}{\sqrt{1-\epsilon}} =
\frac{\sqrt{gH}}{\sqrt{1-A/H}} \approx \sqrt{gH} \bigg{(} 1 +
\frac{A}{2H} \bigg{)}. \label{vel-order1}
\end{equation}
Relations (\ref{eps-order1}) and (\ref{vel-order1}) are well known
from the KdV theory of solitary waves \cite{KdV}, as well as the
shape (\ref{shape-order1}) of the wave, as plotted in Figure
\ref{figure3}.

Unlike the previous one, the second-order approximate equation
(\ref{eqn-order2}) can not be solved in such an elementary way. We
first examine the function $h(x)$ asymptotically, i.e. in the region
where $h(x) \rightarrow 0$. By neglecting nonlinear terms in
equation (\ref{eqn-order2}) we obtain
\begin{equation}
-2 \epsilon \frac{h(x)}{H} + \frac{2}{3} H \cdot h''(x) +
\frac{2}{45} H^3 h^{(4)}(x) = 0, \label{eqn-order2-lin}
\end{equation}
with abbreviation (\ref{eps-def}) used once again. Seeking the
solution in the exponential form $\exp(kx)$ gives four distinct
roots:
\begin{equation}
k_{1,2} = \pm \frac{1}{H} \sqrt{\frac{15}{2}} \cdot \Bigg{(} -1 +
\sqrt{1 + \frac{4}{5} \epsilon} \Bigg{)}^{1/2} \label{k-12}
\end{equation}
and
\begin{equation}
k_{3,4} = \pm \frac{i}{H} \sqrt{\frac{15}{2}} \cdot \Bigg{(} 1 +
\sqrt{1 + \frac{4}{5} \epsilon} \Bigg{)}^{1/2}. \label{k-34}
\end{equation}
The magnitude of these roots is estimated for $\epsilon \ll 1$ to be
\begin{equation}
k_{1,2} \approx \pm \frac{1}{H} \sqrt{3 \epsilon} \quad \textrm{and}
\quad k_{3,4} \approx \pm \frac{i}{H} \sqrt{15}. \label{k-approx}
\end{equation}
While the first two roots are appropriately small, the last two
describe sinusoidal waves with wavelength comparable to the ambient
water height $H$. Accepting such waves would contradict the second
relation of (\ref{term-small2}), which is an assumption leading to
equation (\ref{eqn-order2}). The solitary wave solution therefore
must approach the real exponential functions corresponding to
$k_{1,2}$ as $h(x) \rightarrow 0$: an increasing one ($k_1 > 0$) for
$x \rightarrow -\infty$ and a decreasing one ($k_2 < 0$) for $x
\rightarrow +\infty$.

\begin{figure}[t!]
\centering
\includegraphics[width=6cm]{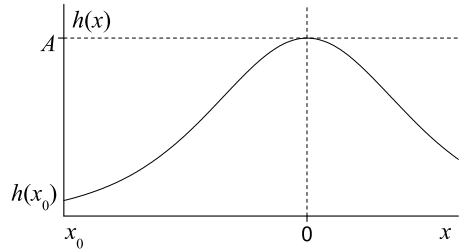}
\caption{Definition of the quantities $h(x_0)$ and $A$, and the
numerical solution of equation (\ref{eqn-order2}) with $x_0$ =
$-$6.78, $h(x_0)$ = 0.01 and $\epsilon$ = 0.1. \label{figure4}}
\end{figure}

In the region between the two limiting cases, we solve equation
(\ref{eqn-order2}) numerically. The initial values for the function
$h(x)$ and its derivatives can be set according to the asymptotic
form described above. If we start from a small value $h(x_0)$
located at a large negative $x_0$, the derivatives have initial
values
\begin{equation}
h^{(n)}(x_0) = h(x_0) \cdot k_1^n. \label{eqn-order2-init}
\end{equation}
If the initial derivatives are not set according to this, short
sinusoidal waves appear, leading to the above-mentioned
contradiction. As long as $h(x_0)$ is sufficiently small, changing
its value only leads to a shift along the $x$-axis. For every
$\epsilon$ therefore we can find a solitary wave solution with a
unique shape and amplitude. The latter is read off as the maximum
value of the numerically obtained function $h(x)$, as shown in
Figure \ref{figure4}. The results obtained for different $\epsilon$
values are summarized in Table \ref{table}, along with two simple
functional forms of $A/H$, the second of which appears to be a
really good approximation of $\epsilon$.

\begin{table}[h]
\begin{indented}
\item[]\begin{tabular}{@{}lllll}
\br
    $\epsilon$  &   $A/H$       &    $A/H - A^2/H^2$      &     $A/H - 21A^2/20H^2$      \\
\mr
    0.01        &   0.0101072   &    0.010005             &     0.0099999                \\
    0.02        &   0.0204382   &    0.020020             &     0.0199996                \\
    0.03        &   0.0310081   &    0.030047             &     0.0299985                \\
    0.04        &   0.0418337   &    0.040084             &     0.0399961                \\
    0.05        &   0.0529342   &    0.050132             &     0.0499921                \\
\br
\end{tabular}
\end{indented}
\caption{The quantity $\epsilon$ versus different functional forms
of $A/H$ obtained from numerical solutions of equation
(\ref{eqn-order2}). \label{table}}
\end{table}

The relationship between $\epsilon$ and the amplitude $A$ indicates
that a correction should be added to relation (\ref{eps-order1}) in
the case of second-order solitary waves:
\begin{equation}
\epsilon = \frac{A}{H} - \frac{21A^2}{20H^2}. \label{eps-order2}
\end{equation}
Although Table \ref{table} only shows values of $\epsilon$ much
smaller than unity, relation (\ref{eps-order2}) remains
approximately valid for larger values. After the critical value of
$\epsilon \approx 0.25$, however, there are no solitary wave
solutions; the function $h(x)$ diverges exponentially. The highest
possible solitary waves are found to have the ratio $A/H \approx
0.5$. While equation (\ref{eqn-order1}) formally allows solitary
waves of arbitrary height, equation (\ref{eqn-order2}) contains a
hint on the experimentally observed instability (wave breaking) of
high solitary waves.

\begin{figure}[t!]
\centering
\includegraphics[width=12cm]{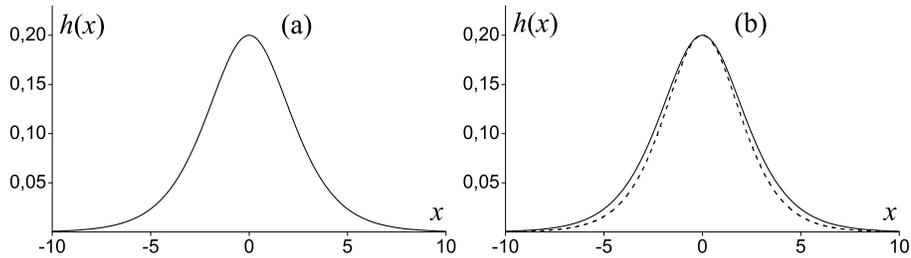}
\caption{Solitary wave of the second-order equation
(\ref{eqn-order2}) obtained numerically with $H$ = 1 and $A$ = 0.2
(a); the same wave plotted together with the analytically known
first-order wave marked by dashed line (b). \label{figure5}}
\end{figure}

The propagation velocity can once again be expressed from equations
(\ref{eps-def}) and (\ref{eps-order2}). This time, however, we have
to keep terms up to the order of $A^2/H^2$ to obtain the velocity
for the solitary waves of the second-order equation
(\ref{eqn-order2}):
\begin{equation}
c = \frac{\sqrt{gH}}{\sqrt{1-\epsilon}} = \frac{\sqrt{gH}}{\sqrt{1 -
A/H + 21A^2/20H^2}} \approx \sqrt{gH} \bigg{(} 1 + \frac{A}{2H} -
\frac{3A^2}{20H^2} \bigg{)}. \label{vel-order2}
\end{equation}
This result is equivalent to that obtained by Laitone with a
systematic expansion of the flow components \cite{Laitone}. Note
that the velocity given by expression (\ref{vel-order2}) is smaller
than the velocity in (\ref{vel-order1}); the maximum correction,
belonging to $A/H \approx 0.5$, is about $(3/80)/(5/4) \approx 3\%$.
The new terms in equation (\ref{eqn-order2}) with respect to
equation (\ref{eqn-order1}) are only small corrections, thus the new
solitary waves plotted in Figure \ref{figure5}(a) are similar to
those of the KdV theory. Comparison in Figure \ref{figure5}(b)
indicates that the second-order solitary waves are slightly longer.

To summarize, the results obtained via the numerical methods show
that the solitary waves of the second-order equation
(\ref{eqn-order2}) are slower and longer than the corresponding KdV
waves.

\section{Velocity measurements} \label{sect-exp}

\begin{figure}[h]
\centering
\includegraphics[width=9cm]{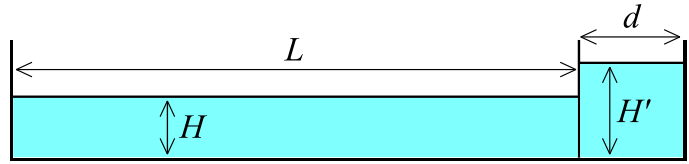}
\caption{The initial set-up used to produce solitary waves ($L$ = 10
m, $d$ = 13 cm, $H$ = 10.3 cm, while $H'$ varied between 17 cm and
24 cm). \label{figure6}}
\end{figure}

In order to experimentally examine the validity of the theory, we
measured the propagation velocity of large amplitude solitary waves
and compared the results with relations (\ref{vel-order1}) and
(\ref{vel-order2}). Solitary waves propagating in a long, narrow
glass tank were observed with a CCD camera standing in the direction
perpendicular to the propagation. The experiments were carried out
with coloured tap water to make the contrast between the fluid and
the background larger. The rightmost part of the tank shown in
Figure \ref{figure6} was separated by a lock-gate and contained
water of height $H'$, larger than the height $H$ of the ambient
fluid. After pulling out the lock-gate, this difference of water
height generated a solitary wave propagating to the left. Solitary
waves of different amplitude were produced by changing the modified
water height $H'$. The camera was placed to look at a region 6 m
away from the right end of the tank to avoid transient phenomena.
Reflected waves propagating to the right were also recorded.

\begin{figure}[h]
\centering
\includegraphics[width=10cm]{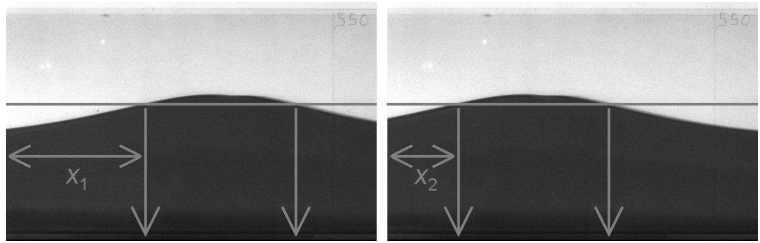}
\caption{Two photographs taken in rapid succession ($\Delta t$ =
0.0635 s) with arrows indicating the methodology for determining the
velocity of the wave: the displacement $\Delta x$ = $x_1 - x_2$ =
7.54 cm gives $c$ = 119 cm s$^{-1}$. \label{figure7}}
\end{figure}

The pictures taken by the camera were afterwards digitally evaluated
by a computer program, which calculated the displacement of the
solitary wave between subsequent images. The profile of the wave
could easily be obtained by finding the border separating the dark
and light regions of the picture. Intersections of the surface and
different horizontal lines were used to monitor the solitary wave on
subsequent images, as shown in Figure \ref{figure7}.

The exact value of the water height $H$ and the amplitude $A$ could
also be read off by the computer program. Both values are obtained
as the difference of appropriate vertical coordinates: the former is
the distance between the bottom of the tank and the surface far in
front of the solitary wave, while the latter is the maximum value of
the surface with respect to $H$. In the case of direct solitary
waves, both values $H$ and $A$ are known with the accuracy of
$\pm$1.5 mm. When examining reflected waves, the maximum value used
to calculate the amplitude $A$ has a larger deviation on subsequent
images, hence the error of $A$ increases to $\pm$2.5 mm.

\begin{figure}[t!]
\centering
\includegraphics[width=10.5cm]{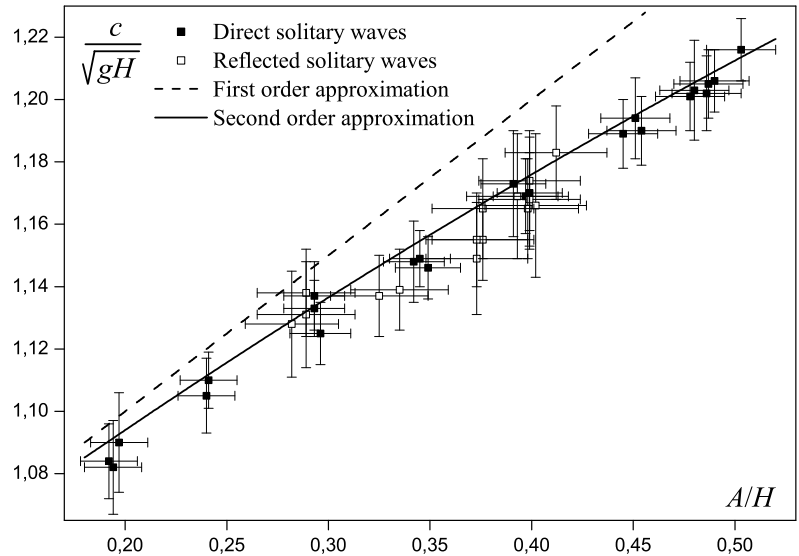}
\caption{Dimensionless propagation velocity of the solitary waves as
a function of $A/H$; the continuous lines mark the theoretical
predictions (\ref{vel-order1}) and (\ref{vel-order2}).
\label{figure8}}
\end{figure}

Since the time interval between two pictures is known with a
negligible relative error of $\sim$ 0.1\%, the velocity can easily
be obtained from the displacement. We used several subsequent images
to fit a line on the relative displacements to reduce the error in
the velocity. This error was estimated as the standard deviation of
the different velocities obtained from tracing the left and right
intersections of 9-26 horizontal lines, their number depending on
the amplitude of the solitary wave. The average values of these
velocities for the same wave are plotted non-dimensionally in Figure
\ref{figure8}, with the vertical error of the points coming from two
main sources. Besides the deviation in the velocity, the error in
the water height $H$ must also be taken into account when
calculating the Froude number
\begin{equation}
\mathrm{Fr} = \frac{c}{\sqrt{gH}}. \label{froude}
\end{equation}
The theoretical predictions (\ref{vel-order1}) and
(\ref{vel-order2}) are also plotted for comparison. Figure
\ref{figure8} clearly shows the coincidence of the measured points
and relation (\ref{vel-order2}), and indicates therefore the
validity of the underlying second-order theory.

\section{Conclusions} \label{sect-conc}

The iteration scheme presented in Section \ref{sect-iter} can be
used to obtain equations describing solitary waves in arbitrary
order of approximation. These equations are ordinary differential
equations, therefore they only give solutions of stationary shape
(one-soliton solutions). By implementing the elementary iteration
scheme of Section \ref{sect-iter}, we obtained the first- and
second-order approximate equations.

The first-order equation was solved in Section \ref{sect-sol} to
recover the well-known solitary waves of the KdV theory. By finding
the solitary wave solutions of the second-order equation and
comparing them to those of the KdV theory, we found that the
second-order solitary waves are slightly slower and longer than the
corresponding KdV waves. The expression obtained for their
propagation velocity is equivalent to the result derived by Laitone
\cite{Laitone}. Velocity measurements of real solitary waves show a
really good correlation between the experimental results and the
second-order theory, which is therefore experimentally verified to
be valid.

It is worth mentioning that further refinements to the theory do not
give observable corrections to the propagation velocity. By taking
the third-order approximation we would obtain $c/\sqrt{gH} = 1 +
A/2H - 3A^2/20H^2 + 3A^3/56H^3$ \cite{Laitone}. In the case of the
highest solitary waves with $A/H \approx 0.5$, the last term is only
responsible for a maximum correction of $\approx 0.5\%$, which is
very difficult to demonstrate in an experimentally reliable way. The
second-order approximation is sufficient to describe shallow-water
solitary waves in all realistic situations.


\ack

We would like to thank Imre M J\'{a}nosi, J W A Robinson and
Tam\'{a}s T\'{e}l for helpful discussions, and Bal\'{a}zs Gy\"{u}re
for technical assistance. This work was supported by the Hungarian
Science Foundation (OTKA) under grant NK72037.


\end{document}